\documentclass[twocolumn]{aastex631}
\usepackage{dcolumn,ulem}
\usepackage{bm}
\usepackage{graphicx}
\usepackage{float}
\usepackage{amsmath}
\usepackage{mathrsfs}
\usepackage{enumerate}
\usepackage[skip=4pt, indent=10pt]{parskip}
\usepackage{orcidlink}
\usepackage{comment}
\usepackage{url}

\newcommand{\fav}{\langle f_{\rm rot} \rangle}

\begin{document}

\title{
Deformation and differential rotation in slowly rotating young intermediate-mass stars
}

\author{Subrata Kumar Panda \orcidlink{0009-0009-1287-6521}}
\affiliation{Department of Astronomy and Astrophysics, Tata Institute of Fundamental Research, Colaba, Mumbai 400005, Maharashtra, India}

\author{Shravan Hanasoge \orcidlink{0000-0003-2896-1471}}
\affiliation{Department of Astronomy and Astrophysics, Tata Institute of Fundamental Research, Colaba, Mumbai 400005, Maharashtra, India}

\begin{abstract}
    Asteroseismology, the study of stellar vibrations, is a method which can probe the structure deformation and internal rotation of stars. Salient among the seismic inferences of rotation from TESS observations are TIC 408165734, whose equatorial rotation rate is 10\% faster than the pole, and TIC 307930890, which has significant radial shear and shows a decreasing spin rate outward through its envelope. We also measure structural deformation in fifteen stars, nine of which are oblate, a finding consistent with expectations for relatively fast-rotating, non-magnetic stars. The difference between polar and equatorial radii in TIC 47639058 is 130 times larger than that for the Sun. The remaining six stars display splittings consistent with a prolate shape (surprisingly), possibly indicating the presence of equatorial toroidal magnetic fields. These inferences provide constraints for numerical simulations and new insights to guide theories of $\delta$ Scuti structure and rotation.
\end{abstract}

\section{Introduction}

Rotation profoundly influences stellar evolution, structure and dynamics. One major effect, especially in fast rotators, is deformation, which turns the stars into ellipsoids, in turn leading to phenomena such as gravity darkening (\citealt{grvdrk, grv_drk}), enhanced accretion efficiencies (\citealt{Stringent}),  anisotropic winds (\citealt{anisotropic_wind}) and mass-loss episodes (\citealt{def_lss}). In binary systems, deformation can impact the tidal interaction, exchange of angular momenta, and energy dissipation (\citealt{crit_rot_accret, tidal}) between the bound objects.

Another important phenomenon known as differential rotation, where different layers of stars spin at varying rates, affects interior mixing processes (\citealt{mixing}) and other dynamics. Differential rotation, as in the case of the Sun, is responsible for the generation of stellar magnetic field through the conversion of kinetic energy into magnetic energy (e.g. \citealt{Spruit1999}). It generates the toroidal component of the stellar magnetic field by winding-up the poloidal components, which results in efficient large-scale magnetic torques and small-scale Lorentz stresses capable of damping the differential rotation (\citealt{Mestel_Weiss, Spruit1999, Spruit_dynamo_dfrot, Mathis_Zahn_2005, Gaurat, Fuller2019}). The interplay between stellar evolution and angular momentum transport mechanisms (\citealt{AMT2019}) collectively determines the core-to-surface rotation gradient, which has been measured in numerous stars using methods of asteroseismology (\citealt{aerts2010asteroseismology, infer_rot, Reeth_2015, rev_fs}).

Non-radial modes of stellar oscillation are used to infer differential rotation and magnetism, because they cause modes of specific angular degree $\ell$ to split into a set of ($2\ell+1$) multiplets. These splittings have enabled the measurements of interior differential rotation (\citealt{Mosser2012}) in 22\% of red giant stars, demonstrating that their cores are spinning faster than the envelopes, but slower than theoretically expected (\citealt{Cantiello_2014, theor_rgb}). Asteroseismic mode splittings also enabled ensemble-scale inference of differential rotation in main-sequence stars (\citealt{rl_str_prlt,  Benomar_2018, Benomar_2023}). Negligible differences between the core and envelope rotation rates of main-sequence stars align well with the expectation that they rotate nearly rigidly due to efficient angular momentum transport (\citealt{Talon_Charbonnel, Eggenberger2022}). These stars rotate slowly and are nearly spherical in shape.

In contrast, hot stars of intermediate masses ($1.5-2.5 M_\odot$) known as $\delta$ Scuti stars, are rapid rotators ($\Omega \sim 10 - 100 \Omega_\odot$) and are consequently deformed - typically oblate in shape due to centrifugal forcing - in structure. Dynamo-generated magnetic fields are thought to be absent in them (\citealt{reviewKepler}). Nevertheless, observations have revealed magnetic fields in some $\delta$ Scuti stars (\citealt{Balona2019, bCas}). The splittings of non-radial modes that provide insights into rotation and deviations are not easily identified in the oscillation spectra of $\delta$ Scuti stars due to challenges in mode identification (\citealt{Nature2020}). Notably in the case of star KIC 11145123, a relatively slowly spinning $\delta$ Scuti star, measurements of internal rotation (\citealt{df-rot-KIC11145123}) and ellipsoidal deformation (\citealt{Gizon_2016}) were possible.

Spectropolarimetric observations have facilitated notable progress in measuring the line-of-sight photospheric magnetic fields in a few $\delta$ Scuti stars. Main-sequence stars of higher mass (O-B types) possess fossil fields in their radiative envelopes, remnants of the dynamo field they generated while being fully convective during their pre-main sequence. In contrast, low-mass main-sequence stars generate magnetic fields through dynamo action in their convective envelopes. $\delta$ Scuti stars (A-F types), having intermediate characteristics, are expected to show magnetic fields of mixed nature -- early (A) type stars having fossil field and later (F) types generating dynamo fields. When it comes to magnetic-field strength, A-type $\delta$ Scuti stars display a phenomenon known as magnetic dichotomy (\citealt{dichotomy1}). Whereas chemically peculiar Ap stars possess fields stronger than 100 gauss, metallic-line Am stars harbour fields weaker than 1 gauss. In the case of Vega, a rapidly spinning non-peculiar A-type star, a 0.6G magnetic field was observed near its pole (\citealt{vega1}).

Thus far, only a few $\delta$ Scuti stars have observationally confirmed detections of magnetic fields. HD 188774, an A-type main-sequence star, was the first confirmed $\delta$ Scuti found to possess fossil magnetic field (\citealt{ds_1_mgntic}), with a magnitude of a few tens of gauss. HD 67523, an F-type chemically peculiar evolved star, was the second $\delta$ Scuti found with a toroidal fossil magnetic field with strength weaker than 1 Gauss (\citealt{ds_2_mgntic}). $\beta$ Cas, an F-type star close to the terminal age main sequence (TAMS), was the third $\delta$ Scuti discovered with poloidal dynamo magnetic field of amplitude on the order of a few tens of Gauss (\citealt{bCas}). Rapid rotation may lead to the thickening of the surface convective layer, thereby favoring the generation of a dynamo field in $\beta$ Cas. HD 41641, an A-type star, was the fourth $\delta$ Scuti detected to possess fossil magnetic field of amplitude $\sim$100G, but with greater topological complexity than the classical dipolar case (\citealt{ds_4_mgntic}). Following these findings, an effort to detect magnetic fields in a potential sample of 12 $\delta$ Scuti stars did not yield positive results (\citealt{B_dlt_str}).

Inferences of deformation and differential rotation across a larger sample of $\delta$ Scuti stars are essential to further our understanding of the dynamics pertinent to this mass range and stellar class. With the availability of a large observational dataset from the TESS mission, and following the discovery of many $\delta$-Scuti pulsators with regularly patterned oscillation (\citealt{Nature2020}), Singh et al. (under review) have identified a group of 38 stars that exhibit rotational splittings. Applying the $a$-coefficient decomposition formalism described in section \ref{sec:theory} to this stellar sample, we measured potential signatures of differential rotation and deformation in 16 stars, whose details are listed in section \ref{sec:result}. Among these, we were only able to constrain the radial and latitudinal differential rotation rates of TIC 307930890 and TIC 408165734, on which we elaborate in sections \ref{sec:rad_diff_rot} and \ref{sec:ldr} respectively. The remaining 14 stars, along with TIC 307930890, exhibit dipole-mode splittings in the form of triplets, including the $m=0$ components. From the asymmetry in the frequency splittings of these stars, we were able to measure their deformations in section \ref{sec:deformn}. We conclude in section \ref{sec:Discussion} by describing how the asymmetric frequency splitting can shed light on the interplay between centrifugal distortion and magnetism in these stars.

\section{Theory of Non-radial mode splitting} \label{sec:theory}

Rotation, centrifugal deformation, and magnetic activity -- collectively break the frequency degeneracy between the multiplets of non-radial stellar oscillation modes (equation \ref{eq:all_split}). The impact of magnetic activity (\citealt{stlr_act}) on oscillation frequency critically depends on the field geometry, and hence cannot be expressed in a closed analytic form. Nevertheless, the Coriolis force due to rotation (\citealt{Hansen1977}) and centrifugal deformation (\citealt{cntrfg_efft}) perturb the degenerate mode frequency in a simpler manner -- as given in equations \ref{eq:del_nu_rot} and \ref{eq:del_nu_cf},
\begin{align} 
    \nu_{\rm n,\ell,m} &= \nu_{\rm n,\ell} + \delta\nu^{\rm rot.} + \delta\nu^{\rm cf.} + \delta\nu^{\rm act.}, \label{eq:all_split} \\
    \delta\nu^{\rm rot} &= (1-C_L) \fav \label{eq:del_nu_rot} \\
    &= (1-C_L) \oint K_{n,\ell} (\vec{r}) f_{\rm rot}(\vec{r}) d^3 \vec{r}, \label{eq:kernel_integrated} \\
    \delta\nu^{\rm cf}  &= \dfrac{2}{3} \dfrac{\ell(\ell+1)-3m^2}{(2\ell-1)(2\ell+3)} \dfrac{\Omega^2 R^3}{GM} \nu_{\rm n,\ell}, \label{eq:del_nu_cf}
\end{align}
where ($M,R,\Omega$) denote the stellar mass, radius, and mean rotation. The Ledoux constant $C_L$ (\citealt{1951Ledoux}) and the rotation-sensitivity kernel ($K_{\rm n,\ell}$) of the oscillation modes are given below in terms of the horizontal and vertical displacements ($\xi_h(\vec{r}), \xi_r(\vec{r})$) and the density profile, $\rho (\vec{r})$.

\begin{align}
    C_L &= \dfrac{\oint \rho ~\xi_h(2\xi_r+\xi_h) d\vec{r}}{\oint \rho ~[\xi_r^2+\ell(\ell+1)\xi_h^2] d\vec{r}} \label{eq:Led}, \\
    K_{n,\ell} &= \dfrac{\rho ~[\xi_r^2 + \ell(\ell+1)\xi_h^2 - 2\xi_r\xi_h - \xi_h^2]}{\oint \rho ~[\xi_r^2 + \ell(\ell+1)\xi_h^2 - 2\xi_r\xi_h - \xi_h^2] d\vec{r}}. \label{Ker}
\end{align}

The Coriolis effect alone does not affect the frequency of the $m=0$ mode - it shifts the $\pm |m|$ modes symmetrically in opposite directions relative to the $m=0$ mode. However, centrifugal distortion and magnetic activity perturb the $m=0$ frequency as well and shift the $\pm |m|$ modes in the same direction (hence asymmetrically with respect to the $m=0$ component). Therefore, the mathematical form of rotational frequency perturbations involve odd-only powers of $m$, while centrifugal deformation and magnetic activity involve even powers of $m$. 

Following notation developed first in helioseismology, when the mode-splitting is projected onto a basis $\mathscr{P}$  (polynomial functions of $m$, \citealt{polyn-zeta}) in form of
\begin{equation} \label{eq:a_cf_exp}
    \nu_{n,\ell,m} = \nu_{n,\ell} + \sum_{j=1}^{j=2\ell} ~a_j(n,\ell) ~\mathscr{P} _j^{(\ell)}(m),
\end{equation}
the mode-splitting decomposes into a series of the so-called $a$-coefficients (\citealt{rot_split_cf}), which are quantities related to perturbative contributions from rotation and differential rotation, stellar deformation, and magnetic field. At leading order, each of these coefficients is independent, allowing us to write a sequence of separate inverse problems. Analytical expressions of a few $\mathscr{P}$ polynomials can be found in \citealt{Benomar_2023} (equation A2-A7).

Rotation-induced perturbations (symmetric in $m$, \citealt{rot_split_cf}) are written in terms of the odd $a$-coefficients (equation \ref{eq:asym_splt}),
\begin{align} \label{eq:asym_splt}
    \delta\nu^{\rm rot} &= \sum_{k=1}^{\ell} ~a_{2k-1} ~\mathscr{P}_{2k-1}^{(\ell)}(m).
\end{align}
The first element of this series, the $a_1$ coefficient, is a measure of mean stellar rotation (\citealt{a1cf_mn_rt}), and it can be estimated from the frequency splittings of modes having $\ell \ge 1$. Similarly, the $a_3$ coefficient probes the extent of latitudinal differential rotation, where spin rates vary across layers from equator to pole, and it can be only determined with modes of degrees $\ell \ge 2$.

Frequency splittings involving centrifugal deformation, low-latitude magnetic activity, tidal deformation, stratification and temperature perturbations (\citealt{Benomar_2023}) are written in terms of even $a$-coefficients (equation \ref{eq:sym_splt}).
\begin{align} \label{eq:sym_splt}
    \delta\nu^{\rm cf,act} &= \sum_{s=1}^{\ell} ~a_{2s}~\mathscr{P}_{2s}^{(\ell)}(m) \\
    a_{2s} &= a_{2s}^{\rm cf} + a_{2s}^{\rm act} + a_{2s}^{\rm tid} + \cdots. \nonumber
\end{align}
The first coefficient of this series, i.e. $a_2$, generally measures the deformation or asphericity (\citealt{Gizon_2016, rl_str_prlt}) of the stars. Combining equation 23 and 7 of \cite{rl_str_prlt}, the relative difference between equatrial and polar radii may be written as
\begin{align} \label{eq:dfm_prime}
    \dfrac{R_{\rm eq}-R_{\rm pole}}{R_{\rm eq}} &= \dfrac{3}{8\pi} \beta_0 = \dfrac{3}{8\pi} \dfrac{\beta_{n,\ell,m}}{Q_{\ell,m} ~\nu_{\rm n,\ell}},
\end{align} 
where, based on their equation 3 and 7, we can substitute
\begin{equation*}
    \beta_{n,\ell,m} = a_2^{n,\ell} \mathscr{P}_2^{(\ell)} =  \dfrac{3m^2-\ell(\ell+1)}{(2\ell-1)} a_2^{n,\ell},
\end{equation*}
and
\begin{equation*}
    Q_{\ell,m} = \dfrac{\ell(\ell+1)-3m^2}{(2\ell-1)(2\ell+3)},
\end{equation*}
to obtain the expression for asphericity in terms of the $a_2$-coefficient and the unperturbed frequency $\nu_{\rm n,\ell}$:
\begin{equation} \label{eq:asphrcty}
    \dfrac{R_{\rm eq}-R_{\rm pole}}{R_{\rm eq}} = -\dfrac{3}{8\pi} \dfrac{(2\ell+3)a_2}{\nu_{\rm n,\ell}}.
\end{equation}

The centrifugal component of $a_2$ (\citealt[equation 12]{Benomar_2023}) may be expressed in terms of mass, radius, stellar rotation and frequency of  the unperturbed central mode (equation \ref{eq:a2_frm_cntrfgl}), 
\begin{equation} \label{eq:a2_frm_cntrfgl}
    a_2^{\rm cf} = -\dfrac{1}{2\ell+3} \dfrac{\Omega^2 R_\star^3}{G M_\star} \nu_{\rm n, \ell}.
\end{equation}

It is straightforward to extract the three leading $a$-coefficients, provided the frequencies of non-radial multiplets are precisely determined (\citealt[eq. A.8, A.9, A.12]{Benomar_2023}),
\begin{align}
    a_1 &= \dfrac{\nu_{\rm n,1,+1}-\nu_{\rm n,1,-1}}{2} \label{eq:a1_drct_fm}, \\
    a_2 &= \dfrac{\dfrac{ \nu_{\rm n,1,+1}+\nu_{\rm n,1,-1} }{2}-\nu_{\rm n,1,0}}{3} \label{eq:a2_direct_fm}, \\
    a_3 &= \dfrac{\dfrac{ \nu_{\rm n,2,+2}-\nu_{\rm n,2,-2} }{4}-\dfrac{  \nu_{\rm n,2,+1}-\nu_{\rm n,2,-1}  }{2}}{5}. \label{eq:a3_direct_fm}
\end{align}
As evident from the above equations, the $a_1$ coefficient effectively measures the half spacing between the $m=\pm1$ components, and hence, the mean rotation. Similarly, the $a_3$ coefficient refers to the deviation between the centroids of $m=\pm1$ and $m=\pm2$ components. The $a_2$ coefficients measures the shifting of the $m=0$ components relative to the centroid of the $m=\pm1$ components. Whereas the odd $a-$coefficients can be inferred even in the absence of the $m=0$ component, measurement of the $a_2$ coefficient requires the presence of $m=0$ mode in the multiplet splitting.

A non-radial mode of degree $\ell$ may be used to determine $a$-coefficients up to $a_{2\ell}$ (equation \ref{eq:a_cf_exp}). Since a majority of the $\delta$ Scuti stars predominantly pulsate in $\ell=$ 0 and 1 modes, it is possible to obtain $a_1$ and $a_2$ for most of them, the latter being measurable only when $m=0$ modes are observable. Occasionally, these stars also exhibit quadrupolar ($\ell=2$) oscillations, in which case it may be possible to retrieve their $a_3$ coefficients.

\section{Result} \label{sec:result}

We applied the theory described in Section \ref{sec:theory} to a group of 16 stars selected from the 38 $\delta$ Scuti stars that exhibit rotational splittings (Singh et al., under review). The stellar structure parameters, e.g., masses, metallicities and ages of these stars, were reported by Singh et al. (under review) along with their average envelope rotation rates ($\fav$). Supplementing these with estimates of the radii obtained from the TESS Input Catalogue (\citealt{Stassun_2019}), we computed the Keplerian break-up rotation rates of these stars (\citealt{Reese2006}) using the equation:
\begin{equation}
    f_K = \dfrac{1}{2\pi} \sqrt{\dfrac{GM}{R^3}}.
\end{equation}
This is the spin rate at which centrifugal force exceeds self-gravity, resulting in increased mass loss.  Keplerian break-up is an important factor that determines the rotation rates for which perturbative theory offers a valid framework for computing oscillation frequencies (\citealt{Reese2006}). \citet{rapid_rotator_2024} classify stars rotating below 10\% of Keplerian critical velocity as slow rotators (see their section 3.3). Similarly, \citet{Ballot} mention that above 15\% of Keplerian breakup rotation, perturbative calculations become inaccurate (see their introduction) at the tolerance level of 0.08$\mu$Hz (\citealt{Lignieres, Reese2006}). 
Because the rotation rates of the stars we analyzed remain below 10\% of their Keplerian break-up rates (Table \ref{tab:star_informn}: column 7), perturbative calculations are likely valid for our case. Compared to $\delta$ Scuti stars in general (\citealt{rot-distribution}), the stars in our sample indeed spin slower, making it easier to identify the rotational splittings present in their oscillation spectra.

\begin{table*}
    \setlength{\tabcolsep}{3.4pt}
    \renewcommand{\arraystretch}{1.4}
    \centering
    \begin{tabular}{ccccccccc} \hline\hline
        Star & $R$ & $M$ & age  & $\fav$ & $f_K$ & $\fav/f_K$ & phenomena \\
         & $(R_\odot)$ & $(M_\odot)$ &(Myr)  & $({\rm d}^{-1})$ & $({\rm d}^{-1})$ & (\%) & \\ \hline
        TIC 307930890 & $1.544 \pm 0.044$ & $1.7 \pm 0.067$ & $17.04 \pm 2.823$  & $0.487 \pm 0.053$ & 5.731 & 8.5 & radial differential rotation,  \\
         &  &  &   &  &  &  & deformation \\
         TIC 14172135 & $1.660 \pm 0.055$ & $1.68 \pm 0.083$ & $14.19 \pm 2.372$  & $0.240 \pm 0.048$ & 5.11 & 4.7 & deformation \\
         TIC 30624832 & $1.536 \pm 0.052$ & $1.6 \pm 0.051$ & $121.06_{-121.06}^{+245.96}$  & $0.351 \pm 0.03$ & 5.603 & 6.3 & deformation \\
         TIC 42827654 & $1.570 \pm 0.098$ & $1.65 \pm 0.054$ & $252.94 \pm 214.818$  & $0.316 \pm 0.01$ & 5.506 & 5.7 & deformation \\
         TIC 59365685 & $1.643 \pm 0.067$ & $1.75 \pm 0.067$ & $286.07 \pm 164.403$  & $0.381 \pm 0.068$ & 5.297 & 7.2 & deformation \\
         TIC 65734585 & $1.584 \pm 0.042$ & $1.64 \pm 0.066$ & $8.78 \pm 2.59$  & $0.501 \pm 0.05$ & 5.417 & 9.2 & deformation \\
        TIC 111840813 & $1.6 \pm 0.05$ & $1.62 \pm 0.053$ & $485.18 \pm 200.07$  & $0.330 \pm 0.053$ & 5.303 & 6.2 & deformation \\
        TIC 165674519 & $1.458 \pm 0.059$ & $1.64 \pm 0.089$ & $708.66 \pm 191.449$  & $0.310 \pm 0.056$ & 6.134 & 5.0 & deformation \\
        TIC 177715827 & $1.458 \pm 0.059$ & $1.6 \pm 0.06$ & $13.65 \pm 3.0$  & $0.436 \pm 0.04$ & 6.059 & 7.2 & deformation \\
        TIC 238641255 & $1.487 \pm 0.038$ & $2.0 \pm 0.092$ & $174.75 \pm 152.732$  & $0.481 \pm 0.108$ & 6.577 & 7.3 & deformation \\
        TIC 365852391 & $1.535 \pm 0.054$ & $1.56 \pm 0.06$ & $656.83 \pm 205.925$  & $0.232 \pm 0.019$ & 5.538 & 3.5 & deformation \\
        TIC 377257563 & $1.630 \pm 0.046$ & $1.6 \pm 0.075$ & $8.94 \pm 2.411$  & $0.438 \pm 0.005$ & 5.125 & 7.1 & deformation \\
        TIC 405483817 & $1.654 \pm 0.099$ & $1.68 \pm 0.091$ & $8.54 \pm 2.518$  & $0.338 \pm 0.026$ & 5.138 & 6.5 & deformation \\
        TIC 423159418 & $1.496 \pm 0.047$ & $1.64 \pm 0.06$ & $8.92 \pm 2.955$  & $0.463 \pm 0.039$ & 5.902 & 7.8 & deformation \\
         TIC 47639058 & $2.507 \pm 0.125$ & $--$ & $--$  & $--$ & $--$ & $--$ & deformation \\
        TIC 408165734 & $1.609 \pm 0.056$ & $1.68 \pm 0.092$ & $7.93 \pm 2.76$  & $0.3 \pm 0.051$ & 5.355 & 5.6 & latitudinal differential rotation \\ \hline
    \end{tabular}
    \caption{
    Structure and rotational properties of the 16 stars analyzed for this work. The columns from left to right refer to stellar radii (from TESS Input Catalogue), masses, ages, and mean envelope rotation rates (from Singh et al., under review), Keplerian break-up rates ($f_K = \sqrt{GM/R^3}/2\pi$), stellar rotation rates as a fraction of $f_K$, and the characteristics we inferred from the measured frequency splittings.
    }
    \label{tab:star_informn}
\end{table*}

\begin{figure*}
    \centering
    \includegraphics[width=1\linewidth]{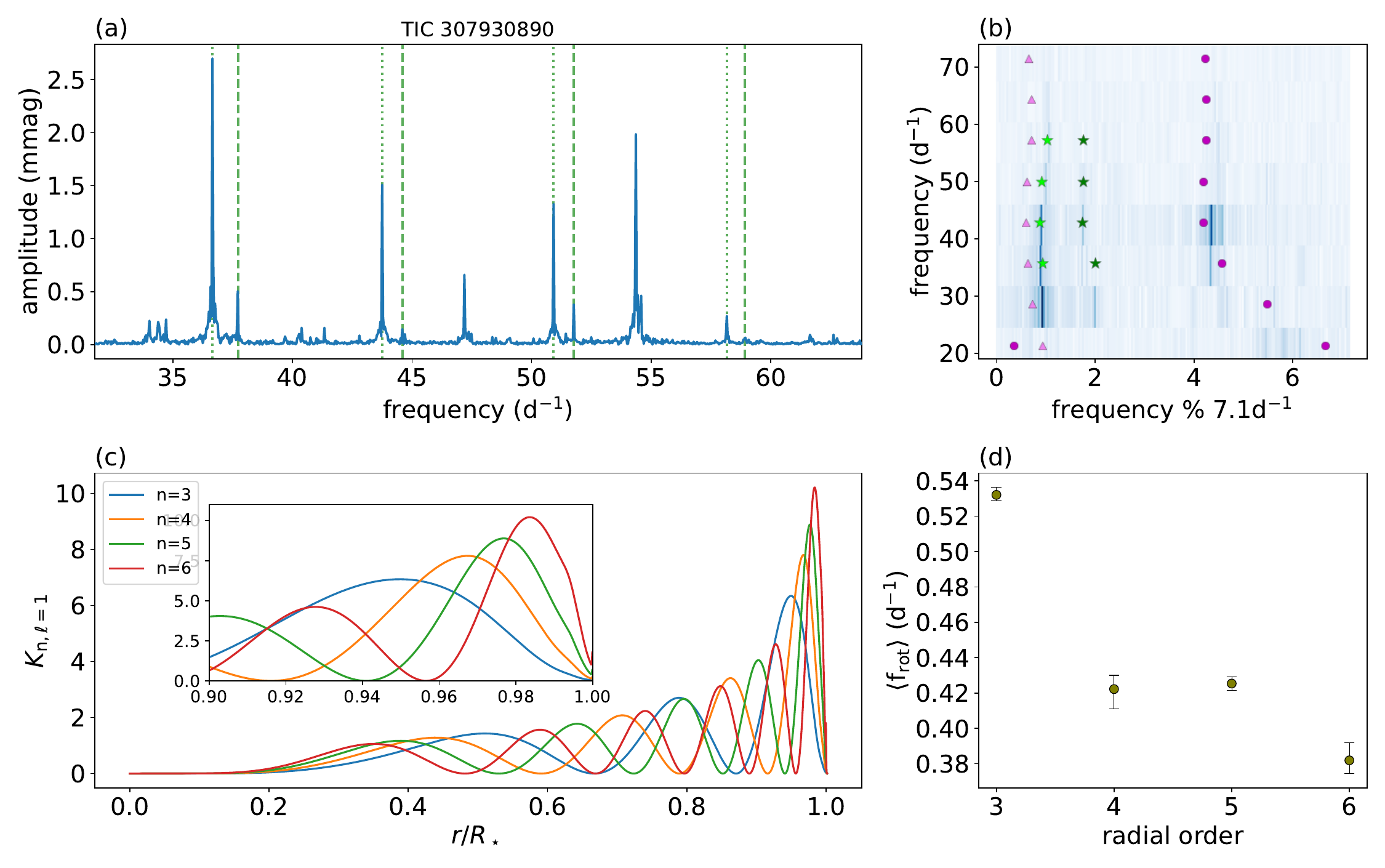}
    \caption{
    Radial differential rotation in a $\delta$ Scuti star.
    (a) Power spectrum of TIC 307930890 with rotationally split dipole-mode ($\ell=1$) doublets at four radial orders. The $m=-1$ ($+1$) components are marked with dotted (dashed) lines.
    (b) The \'echelle diagram associated with this spectrum, obtained by vertically stacking equal-width segments of oscillation spectra, which aligns the modes of a given harmonic degree ($\ell$) into separate vertical ridges. The ridge on the right comprises $\ell=0$ modes. The ridge on the left, appearing to be split into two columns, correspond to the $m=-1$ and $m=+1$ components of the dipole ($\ell=1$) modes. The dipole doublets have been marked with stars symbols, while the circle and triangle symbols anchor the best fit model's radial and dipole modes respectively.
    (c) Radial variation of rotation kernels of the four dipole modes. The kernel represents the modes' ability to sense stellar rotation at specific regions. The inset zooms into the outer 10\% portion of the star, which shows that, lower the radial order, the deeper it probes.
    (d) kernel-weighted (envelope) rotation rates obtained for the four dipole modes as a function of their radial order ($n$).
    }
    \label{fig:rad_df_rot1}
\end{figure*}

\subsection{Radial Differential Rotation} \label{sec:rad_diff_rot}
Since stellar cores shrink and envelopes swell during their evolution, the moments of inertia associated with the cores decline, causing them to spin faster than the envelopes. This leads to the emergence of a rotation gradient along the radial direction.
Resonant acoustic waves (p-modes) propagate in different geometric cavities within stars and by combining all the information they contain, it may be possible to infer details of the internal rotation profiles. Depending on the rotation rates at the regions where the mode eigenfunctions possess the largest amplitudes, the observed mode splittings can vary. Hence, the mean rotation rates $\fav$ obtained from dipole-mode splitting generally varies with their radial orders. The coefficient $a_1$ serves as an indicator of this mean rotation, and may be obtained from the half-spacing between the $m=\pm 1$ modes (equation \ref{eq:a1_drct_fm}).

TIC 307930890 is the best candidate among our sample to demonstrate radial differential rotation, since it exhibits a continuous sequence of four rotationally split dipole doublets, whose frequency spacings gradually decrease across the radial orders (Fig. \ref{fig:rad_df_rot1}a). The apparent absence of the $m=0$ components at all radial orders suggests either a high inclination angle or the involvement of a systematic mode-suppression mechanism in this star. It offers an ideal opportunity to probe the differential rotation profiles in the outer envelope of this star since the acoustic modes are predominantly sensitive to the near-surface regions of $\delta$ Scuti stars.

\subsubsection{Measuring average rotational splitting} \label{sec:av_rot_splitting}

\begin{figure*}
    \centering
    \includegraphics[width=\linewidth]{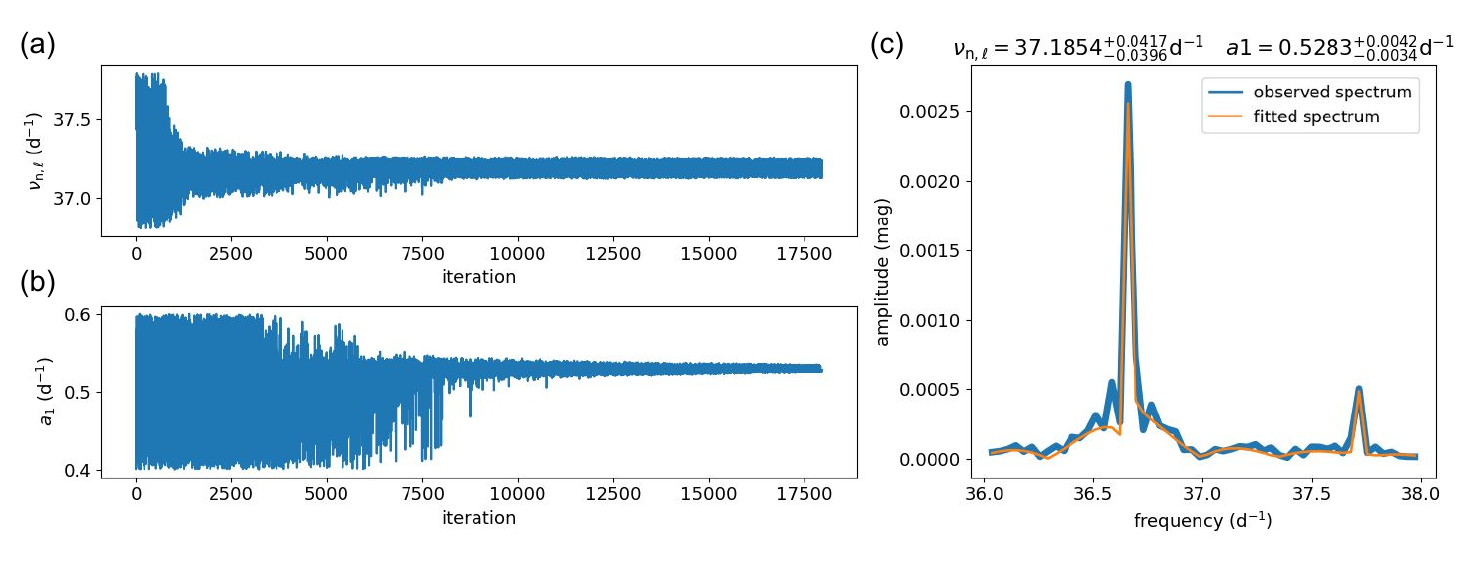}
    \caption{
    Fitting the first mode splitting of TIC 307930890.
    (a) Sequences of parameter $\nu_{\rm n,1}$, the unperturbed mode frequency, as traced by the sampler with subsequent iterations. The sampler, after wandering through the parameter space over thousands of epochs, eventually converges, allowing estimation of the best-fit parameters with uncertainty.
    (b) The similarity trace plot for parameter $a_1$, which relates to the mean rotation rate corresponding to the specific radial order.
    (c) Comparing the observed spectrum of the multiplet and the spectrum modeled using the best-fit parameters shown in the figure title along with their $1-\sigma$ uncertainties.
    }{\bf (A set of 4 figures is available in the online journal.)}
    \label{fig:figset1_1}
\end{figure*}

Expanding equation \ref{eq:a_cf_exp} for $\ell=1$, the frequency splittings of dipole modes can be written in the form
\begin{equation} \label{eq:a_exp_for_l1}
    \nu_{n,1,m} = \nu_{n,1} + a_1 m + a_2 (3m^2-2),
\end{equation}
of which we are interested in the rotation term, $a_1$. This coefficients effectively measure the mean rotation rates up to an additional factor involving the effect of Coriolis force, i.e., $(1-C_L)\fav$. We used the Python package \texttt{Dynesty} (\citealt{Speagle_2020, sergey_koposov_2024_12537467}) to retrieve the probability distributions of the $a_1$ coefficients by individually fitting the oscillation spectra of the dipole doublets at each radial order.

The power spectral profile of each multiplet may be characterized using a parameter set $\theta$, comprising 7 quantities: the unperturbed frequency $\nu_{\rm n,1}$, the  \{$a_1, a_2$\} coefficients, and heights ($h_m$) and widths ($w_m$) of the $m = \pm 1$ components. Because the $m=0$ components are not visible, their heights and widths are redundant for fitting the spectra. Thus, we did not include them in the definition of $\theta$. The inference of $a_1$ is not affected by the absence of the $m=0$ component, as described earlier. Given these parameters, the power spectrum  as a function of frequency $\nu$ of a multiplet may be constructed using the sinc profile
\begin{equation} \label{eq:sinc}
    S_{\rm model}(\theta) = \sum_{m=-\ell}^{\ell} h_m ~{\rm sinc} \left(\dfrac{\nu - \nu_{\rm n,\ell,m}}{w_m/2}\right),
\end{equation}
where the eigen-frequencies, i.e., $\nu_{\rm n,\ell,m}$, are generated using the equation \ref{eq:a_exp_for_l1}.

For a particular rotationally split multiplet, we assigned each 7 parameter a uniform prior
\begin{equation}
    P(\theta) = U(\theta;a,b) \label{eq:prior}
\end{equation}
across appropriate interval $[a-b]$, compatible with the roughly estimated initial guesses from the power spectra. A Gaussian likelihood function
\begin{equation}
    \mathcal{L}(S_{\rm obs}|\theta) = e^{-\dfrac{1}{2} \left( \dfrac{S_{\rm obs}-S_{\rm model}(\theta)}{\sigma} \right)^2}\label{eq:likelihood}
\end{equation}
was used to determine the degree of similarity between the spectrum $S_{\rm model}(\theta)$ generated by the parameter set $\theta$ (equation \ref{eq:sinc}) and the observed spectrum $S_{\rm obs}$, within the tolerance given by the noise level $\sigma$ determined as the median of the observed spectrum. The posterior distribution
\begin{equation}
    P(\theta|S_{\rm obs}) = \mathcal{L}(S_{\rm obs}|\theta) ~P(\theta) \label{eq:pst}
\end{equation}
of the parameter set is obtained by multiplying the likelihood with the prior distribution. The algorithm samples the parameter space by performing random walks based on the posterior probability calculated after each step and may eventually converge to narrowing parameter domains after multiple iterations. We traced all instances of the parameters sampled and discarded the first half of the sequence as the sampler is yet to converge further. From the parameters sampled in the last half iterations, we calculated median, $16^{\rm th}$ and $84^{\rm th}$ percentile to determine the best fit parameters, along with the $1-\sigma$ uncertainties.

We repeated the above procedure for each individual multiplet and obtained the posterior distributions for all the parameters, of which most important are the $\nu_{\rm n,1}$ and the $a_1$ coefficients. We show the results from fitting the first multiplet in figure \ref{fig:figset1_1}, demonstrating the trace plots of the sampled parameters $a_1$ and $\nu_{\rm n,1}$, and comparing the model spectrum generated from the best-fit parameters against the observed spectrum.

The best-fit model for TIC 307930890 (Singh et al., in preparation) suggests this star has a mass of $1.7M_\odot$, metallicity $Z=0.018$, and an age of $17$ Myr. Simulating a stellar model with these properties using \texttt{MESA} (\citealt{MESA2011, MESA2013, MESA2015, MESA2018, Paxton2019}) and performing a pulsation calculation with \texttt{GYRE} (\citealt{GYRE2013, GYRE2017, GYRE2020}), we noticed the $\ell=1$ frequencies of radial orders between 3 and 6 appear in the range of dipole modes observed in this star. Thus, we calculated the Ledoux constants $C_L$ (equation \ref{eq:Led}) for these modes, and divided the $a_1$ coefficients by the factor $(1-C_L)$ to retrieve rotation rates $\fav$ (equation \ref{eq:del_nu_rot}). The fitted $a_1$ coefficients for all radial orders, their Ledoux constant $C_L$, and accordingly extracted $\fav$ values are given in Table \ref{tab:a1_CL_fav}.

\begin{table}
    \centering
    \setlength{\tabcolsep}{12.4pt}
    \renewcommand{\arraystretch}{1.4}
    \begin{tabular}{cccc} \\ \hline\hline
        $n$ & $a_1$ & $C_L$ & $\fav$ \\ 
            & (${\rm d}^{-1}$) & & (${\rm d}^{-1}$) \\ \hline
        3 & $0.5283_{-0.0034}^{+0.0042}$ & 0.0071 & $0.5320_{-0.0034}^{+0.0042}$ \\
        4 & $0.4182_{-0.0112}^{+0.0077}$ & 0.0094 & $0.4222_{-0.0113}^{+0.0077}$ \\
        5 & $0.4206_{-0.0039}^{+0.0037}$ & 0.0111 & $0.4253_{-0.0039}^{+0.0037}$ \\
        6 & $0.3774_{-0.0074}^{+0.0099}$ & 0.0118 & $0.3819_{-0.0074}^{+0.0100}$ \\ \hline 
    \end{tabular}
    \caption{
    Asteroseismic characteristics of the four rotationally split modes observed in TIC 307930890. The columns from left to right provide the radial orders of the splittings, the derived $a_1$ coefficients, the structure-dependent Ledoux constants, and the mean rotation rates obtained by dividing the $a_1$ coefficients by the $(1-C_L)$ factor.
    }
    \label{tab:a1_CL_fav}
\end{table}

We demonstrate the radial order ($n$) dependence of $\fav$ in Fig. \ref{fig:rad_df_rot1}(d), which depicts its descending nature with increasing radial order. Rotation kernels, which capture the sensitivities of the acoustic modes to the rotation rate as a function of radius, peak in the outermost layers of the envelope, where the modes resonate strongly (Fig. \ref{fig:rad_df_rot1}c). The sensitivity gradually increases toward the surface layers with growing radial order (see the zoomed inset). Thus, we expect an outwardly decreasing spin rate for this star.

\subsubsection{Inference of spatial rotation profile} \label{sec:rot_coordinate}

The mean rotation rates $\fav$ obtained above for the 4 radial orders represent the kernel-weighted averages of the depth-dependent rotation profile $f_{\rm rot} (r)$ (equation \ref{eq:kernel_integrated}). Detailed investigation of differential rotation requires inference of the spatially varying rate of rotation $f_{\rm rot} (r)$, which is usually obtained by inverting the kernel-integrated expression for rotational splittings $\fav$. Inversions can work well provided enough rotational splittings are available in the spectra (\citealt{inversion_2016}). Given that we only have four rotational splitting measurements, we are unable to apply a similar approach to infer rotation rates at high spatial resolution. Because of this limitation of the inverse problem, we resorted to a forward-modeling approach where we optimized the rotation rates at a few arbitrarily chosen positions so as to fit the kernel-weighted rotational splittings observed for the four radial orders.

For the specific stellar model we analyzed here, all kernels appear to peak within the outer 10\% of the envelope (Fig. \ref{fig:rad_df_rot1}c). Thus, we sought to obtain the rotation profile on a coarse grid comprising 3 radial depths, i.e., $r = \{0.91, ~0.95, ~0.99\} R_\star$, having grid resolution $\Delta r = 0.04R_\star$. We discretized the integral expression (equation \ref{eq:kernel_integrated}) for $\fav$  on this grid in form of
\begin{equation} \label{eq:rot_splt_dscrt}
    \fav_{n,\ell} = \sum_{i=1}^{N_{\rm grid}} K_{\rm n,\ell}(r_i) f_{\rm rot}(r_i) \Delta r,
\end{equation}
with the aim of optimizing the rotation rates $f_{\rm rot}(r^i)$ at the 3 radial positions mentioned above, using the same \texttt{Dynesty} sampler. Solving four linear equations comprising three independent degrees of freedom ensures the uniqueness of the obtained solution.

We chose a uniform prior
\begin{equation}
    P(\Omega) = U(\Omega;a,b) \label{eq:rot_fit_lim}
\end{equation}
between appropriate ranges $[a-b]$ for each of the three rotation parameters, being compatible with the individual $a_1$ coefficients. Given $\Omega$, the set of 3 parameters \{$f_{\rm rot,r_1}, f_{\rm rot,r_2}, f_{\rm rot,r_3}$\}, the rotational splitting ($\delta\nu_j^{\rm model}$) at different radial orders $j$ were obtained using equation \ref{eq:del_nu_rot}. The probabilities of these parameters regenerating the observed mode-splittings were calculated using 
\begin{equation}
    \mathcal{L}(\delta\nu^{\rm obs}|\Omega) = e^{-\dfrac{1}{2} \sum_{j=3}^6 \left( \dfrac{\delta\nu_j^{\rm obs}-\delta\nu_j^{\rm model}(\Omega)}{\sigma_j^{\rm obs}} \right)^2},\label{eq:rot_fit_likelihood}
\end{equation}
the Gaussian likelihood function of the differences between the observed and modeled splittings scaled by the uncertainties $\sigma_j^{\rm obs}$, taken as the frequency resolution of the spectrum. \texttt{Dynesty} samples the three-dimensional \{$\Omega$\} space following the posterior probability
\begin{equation}
    P(\Omega|\delta\nu^{\rm obs}) = \mathcal{L}(\delta\nu^{\rm obs}|\Omega) ~P(\Omega) \label{eq:rot_fit_pst}
\end{equation}
obtained from the product of the likelihood and the prior functions. The statistics of the parameters traced by the sampler indicates how likely they fit the observed rotational splittings.

\begin{figure*}
    \centering
    \includegraphics[width=\linewidth]{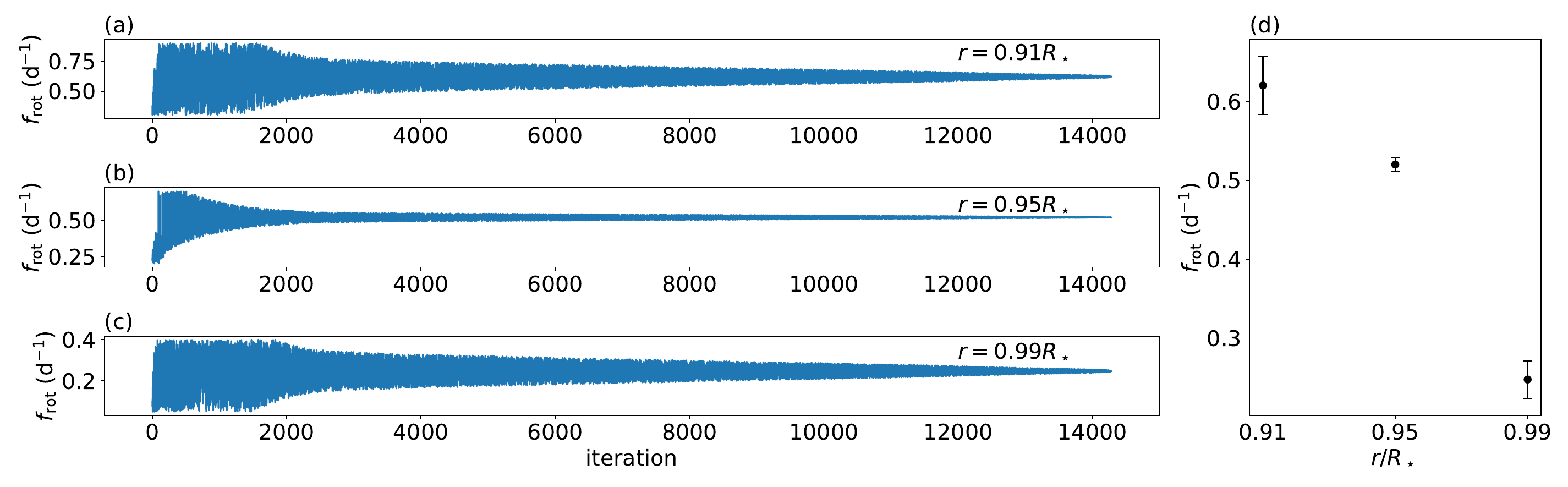}
    \caption{
    Optimizing the spatial rotation of TIC 307930890 over a three zone grid within its outer envelope. (a,b,c) Trace plots of the rotation rates at $r=0.91, 0.95, 0.99 R_\star$, as a function of iterations made by the sampler. (d) Best-fit rotation rates and uncertainties shown over the radial coordinates of the assumed spatial grid.
    }
    \label{fig:rad_df_rot3}
\end{figure*}

In figure \ref{fig:rad_df_rot3}, we demonstrate how the rotation rates were sampled by the algorithm with successive epochs. We rejected the first half of the iterations of these sequences as the sampler converges after traversing the parameter space for many epochs. For the parameters sampled in the final set of iterations, we computed the median and the $16^{\rm th}$, and $84^{\rm th}$ percentiles to obtain the best-fit rotation rates and the associated $1-\sigma$ uncertainties. The rotation rates of the best-fit model obtained at the aforementioned positions are shown in Figure \ref{fig:rad_df_rot3}(d), where the spin rates appear to decline radially outwards, i.e. from $0.62 {\rm d}^{-1}$ at $r=0.91R_\star$ to $0.24 {\rm d}^{-1}$ at $r=0.99R_\star$. Almost 60\% drop in the rotation rate over a depth equal to 8\% of the stellar radius, indicates the possible existence of a strong radial differential rotation in this star.

\subsection{Deformation in Shape} \label{sec:deformn}

\begin{figure*}
    \centering
    \includegraphics[width=\linewidth]{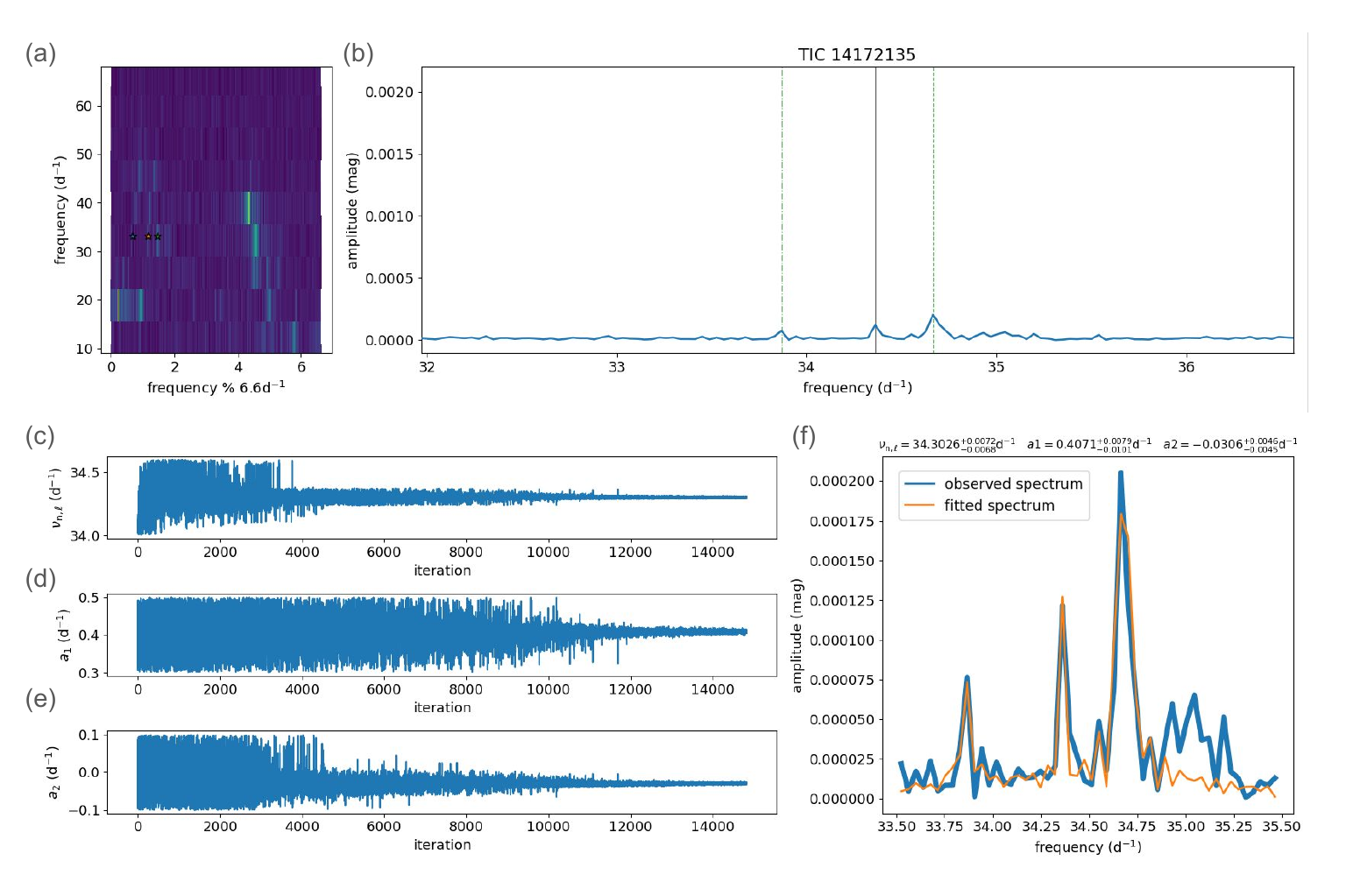}
    \caption{
    Analyzing the asymmetric dipole mode splitting of TIC 14172135.
    (a) \'echelle diagram of pulsation spectrum of the star, with the star symbols marking the dipole triplet we used for calculation of the $a_2$ coefficients.
    (b) A zoomed-in power spectrum of the star, with vertical lines annotating the modes selected from the \'echelle diagram.
    (c,d,e) Trace plots of the unperturbed frequency $\nu_{\rm n,1}$, $a_1$ and $a_2$ coefficients sampled by the algorithm, demonstrating gradual convergence with increasing iterations.
    (f) Comparison between observed spectrum and the  generated analog using the best-fit asteroseismic parameters (shown in the figure title) obtained form the last half of the iterations of the sampler.
    }{\bf (A set of 15 figures is available in the online journal.)}
    \label{fig:figset2_1}
\end{figure*}

In the case of a slowly rotating star lacking significant magnetic field, the non-radial multiplets are nearly equally spaced in frequency space 
However, at moderate-to-fast rotations, stars deform more significantly into ellipsoids, causing the frequency spacing between adjacent azimuthal components to no longer remain constant. For the same reason, the $m=0$ mode in some stars does not appear exactly mid-way between the $m=\pm 1$ components. The position of the $m=0$ mode relative to the $m=\pm 1$ doublets is sensitive to stellar distortion. Even slight distortions in distant stars contain seismic imprints, allowing us to draw inferences (\citealt{Gizon_2016}).

As in equation \ref{eq:a_exp_for_l1}, rotation and magnetism shift the $m=0$ mode by the amount $-2a_2$ relative to the central frequency $\nu_{\rm n,\ell}$, while the $m=\pm 1$ components are displaced by $a_2 \pm a_1$ ($\approx \pm a_1$, since $|a_2| \ll a_1$).  The coefficient $a_2$ captures the asymmetry in the position of the $m=0$ component relative to the mean of the $m=\pm1$ doublets (equation \ref{eq:a2_direct_fm}). Hence, $a_2<0$ (oblate) for stars with the $m=0$ component closer to the $m=+1$ mode and $a_2>0$ (prolate) for stars where the $m=0$ peak is proximal to the $m=-1$ mode.
Therefore, the position of the $m=0$ mode serves as a gauge with which to infer distortions in stellar structure.

The observationally measured $a_2$ coefficient contains additive contributions from centrifugal distortion and equatorial magnetic fields. The centrifugal component of $a_2$ is always a negative quantity (equation \ref{eq:a2_frm_cntrfgl}), whereas the contribution from magnetic activity can assume positive values (\citealt[equation 14]{Benomar_2023}). Thus positive $a_2$ values are suggestive of magnetic activity in the star, substantial enough to dominate the centrifugal effect. Cool stars, characterized by thick convective envelopes, generate dynamo magnetic fields and frequently exhibit such activity.

While centrifugal force and tidal interactions cause equatorial-bulging in stars, making them oblate, toroidal magnetic field can counteract this effect turning them prolate (\citealt{dfm_prlt}). Precise frequency analysis suggests a prolate shape for 16 Cyg A (\citealt{rl_str_prlt}). KIC 11145123, a $\gamma$ Dor-$\delta$ Scuti hybrid, has been observed to be less oblate than expected from its centrifugal deformation alone, thereby indicating the presence of a near-equatorial magnetic field (\citealt{Gizon_2016}).

Among the sample of 38 stars, 15 had visible $m=0$ modes positioned between the $m=\pm 1$ multiplets. One such triplet splitting and related \'echelle diagram of a star are shown in Fig.~\ref{fig:figset2_1}. We attempted to determine the $a_2$ coefficients (a proxy of shape distortion) for these stars by fitting their split modes with equation \ref{eq:a_exp_for_l1}. Although the procedure is very similar to that in section \ref{sec:av_rot_splitting}, here we modeled the split triplets with 9 free parameters: the unperturbed frequency $\nu_{\rm n,1}$, $a_1$ and $a_2$ coefficients, heights ($h_m$) and widths ($w_m$) of the $m=\{\pm1,0\}$ components. We used the \texttt{Dynesty} sampler to draw sequences of these parameters according to their posterior distributions obtained from the products of uniform priors and Gaussian likelihoods between observed and modeled power spectra. Discarding the first half of the iterations of the sampled parameters, we calculated the median of the remaining iterations to estimate best-fit solutions and $16^{\rm th}(84^{\rm th})$ percentiles to compute the $1\sigma$ uncertainties.
The $a_1$ and $a_2$ coefficients calculated in this way for the 15 stars are given in Table \ref{tab:a1a2}. Our sample includes 9 stars with $a_2<0$ (oblate) and 6 stars with $a_2>0$ (prolate), however, with $1\sigma$ (thus, $\sim 68\%$) confidence.

\begin{figure}
    \centering
    \includegraphics[width=\linewidth]{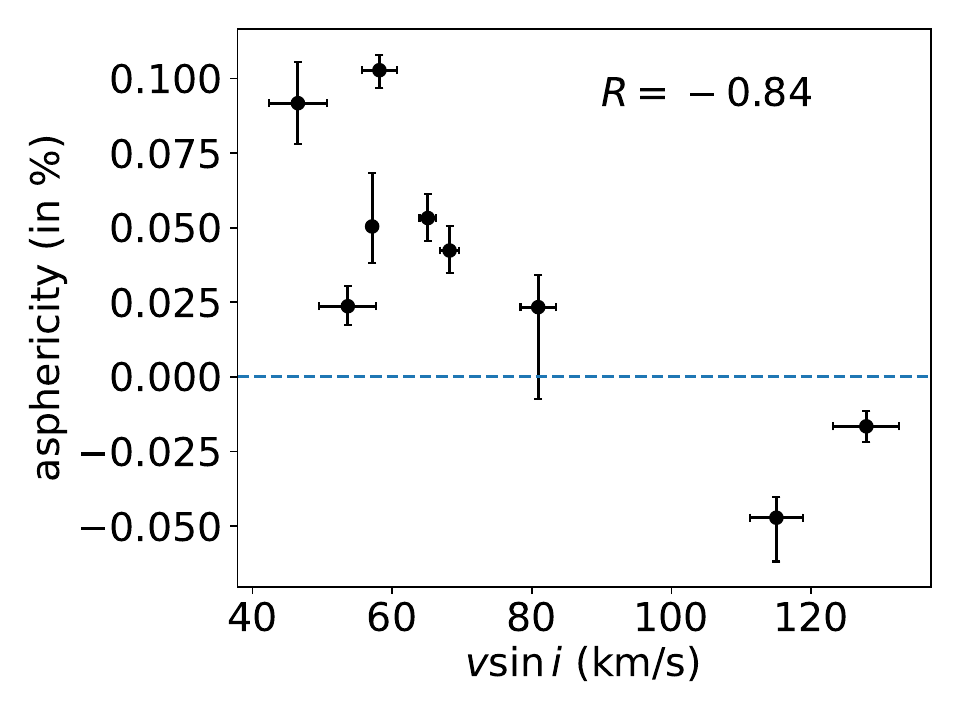}
    \caption{
    Asphericity, the relative difference between equatorial and polar radii, plotted against the line-of-sight projected rotational velocity ($v \sin i$). The latter was available in literature for only 9 stars of our sample.
    }
    \label{fig:asphr_vsini}
\end{figure}

The deviation from sphericity for these stars, defined as $1 - R_{\rm pole}/R_{\rm eq}$ and calculated using equation \ref{eq:asphrcty}, are given in Table \ref{tab:a1a2}, which range from -0.05\% (prolate) to +0.10\% (oblate). These distortions are approximately $\sim 100$ times larger than that of KIC 11145123, a slowly rotating A type star, whose asphericity was measured to be $(1.8\pm0.6)\times 10^{-4}\%$ (\citealt{Gizon_2016}). Deformations measured for the stars in our sample exceed $\sim 100$ times the Solar oblateness $(8.19 \pm 0.33)\times 10^{-4}$\% (\citealt{Meftah2015}).

While neither $a_2$ nor the asphericity showed any strong dependence on other observables, they exhibited a faint correlation with the spectroscopically measured $v\sin i$ (Fig. \ref{fig:asphr_vsini}) -- the latter being available for only 9 of the 15 stars (\citealt{gaiadr3}). Pearson correlation coefficient $R$, which measures the linear correlation between these two quantities, is found to be $-0.84$ in this case. Asphericity takes positive values for oblate stars (equation \ref{eq:asphrcty}), whereas for prolate objects it turns out negative. With the limited data presented here, stars appear to become prolate at larger rotation rates -- which may be linked to the  generation of strong equatorial magnetic fields, capable of counteracting centrifugal deformation.

\begin{table*}
    \setlength{\tabcolsep}{7pt}
    \renewcommand{\arraystretch}{1.4}
    \centering
    \begin{tabular}{cccccccc}
    \hline \hline Star & $\Delta\nu ~\rm (d^{-1})$ & $a_1 ~\rm (d^{-1})$ & $a_2 ~\rm (d^{-1})$ & $\dfrac{R_{\rm eq}-R_{\rm pol}}{R_{\rm eq}} ~(\%)$ & ${\rm v \sin i}$ (km s$^{-1}$) \ \\ \hline \
TIC  14172135  &  $6.426 \pm 0.328$  &  $0.4071_{-0.0101}^{+0.0079}$  &  $-0.0306_{-0.0045}^{+0.0046}$  &  $+0.0532_{-0.0078}^{+0.0080}$ &  $65.116 \pm 1.191$ \\
TIC  30624832  &  $6.659 \pm 0.221$  &  $0.4149_{-0.0262}^{+0.1074}$  &  $+0.0174_{-0.0321}^{+0.0112}$  &  $-0.0485_{-0.0896}^{+0.0312}$ & \\
TIC  42827654  &  $7.061 \pm 0.540$  &  $0.3086_{-0.0037}^{+0.0046}$  &  $-0.0260_{-0.0046}^{+0.0051}$  &  $+0.0423_{-0.0075}^{+0.0083}$ & $68.227 \pm 1.364$ \\          
TIC  59365685  &  $6.731 \pm 0.252$  &  $0.3309_{-0.0094}^{+0.0094}$  &  $-0.0141_{-0.0038}^{+0.0040}$  &  $+0.0236_{-0.0063}^{+0.0067}$ & $53.657 \pm 4.080$ \\
TIC  65734585  &  $6.889 \pm 0.298$  &  $0.4126_{-0.0163}^{+0.0199}$  &  $+0.0206_{-0.0060}^{+0.0078}$  &  $-0.0346_{-0.0101}^{+0.0131}$ & \\
TIC 111840813  &  $6.583 \pm 0.325$  &  $0.3387_{-0.0084}^{+0.0055}$  &  $-0.0287_{-0.0070}^{+0.0102}$  &  $+0.0504_{-0.0123}^{+0.0179}$ &  $57.139 \pm 0.670$ \\
TIC 165674519  &  $6.629 \pm 0.334$  &  $0.3549_{-0.0660}^{+0.0200}$  &  $-0.0162_{-0.0213}^{+0.0074}$  &  $+0.0233_{-0.0307}^{+0.0106}$ &  $80.922 \pm 2.532$ \\
TIC 177715827  &  $7.500 \pm 0.226$  &  $0.4364_{-0.0189}^{+0.0165}$  &  $+0.0153_{-0.0064}^{+0.0068}$  &  $-0.0170_{-0.0071}^{+0.0075}$ & \\
TIC 238641255  &  $7.174 \pm 0.293$  &  $0.5505_{-0.0109}^{+0.0100}$  &  $+0.0142_{-0.0046}^{+0.0044}$  &  $-0.0165_{-0.0053}^{+0.0051}$ &  $127.918 \pm 4.714$ \\
TIC 307930890  &  $7.212 \pm 0.268$  &  $0.4207_{-0.0039}^{+0.0034}$  &  $-0.0335_{-0.0288}^{+0.0371}$  &  $+0.0389_{-0.0334}^{+0.0431}$ & \\
TIC 365852391  &  $6.867 \pm 0.206$  &  $0.4554_{-0.0374}^{+0.0132}$  &  $+0.0393_{-0.0122}^{+0.0057}$  &  $-0.0472_{-0.0146}^{+0.0068}$ &  $115.039 \pm 3.771$ \\
TIC 377257563  &  $5.906 \pm 0.367$  &  $0.4470_{-0.0096}^{+0.0082}$  &  $-0.0386_{-0.0058}^{+0.0058}$  &  $+0.0917_{-0.0138}^{+0.0138}$ &  $46.508 \pm 4.124$ \\
TIC 405483817  &  $6.980 \pm 0.228$  &  $0.2964_{-0.0858}^{+0.0130}$  &  $-0.0323_{-0.0110}^{+0.0083}$  &  $+0.0540_{-0.0185}^{+0.0139}$ & \\
TIC 423159418  &  $6.837 \pm 0.270$  &  $0.4470_{-0.0530}^{+0.0172}$  &  $+0.0086_{-0.0173}^{+0.0074}$  &  $-0.0119_{-0.0240}^{+0.0102}$ & \\
TIC 47639058   &  $4.880 \pm 0.200$  &  $0.4417_{-0.0067}^{+0.0097}$  &  $-0.0689_{-0.0040}^{+0.0035}$  &  $+0.1028_{-0.0059}^{+0.0052}$ &  $58.183 \pm 2.509$ \\ 
Sun     & $11.681 \pm 0.021$  &  $0.0363\pm0.0008$  &  &  $+0.0008\pm0.00003$ &  $1.971\pm0.003$ \\ \hline
    \end{tabular}
    \caption{Large frequency separation ($\Delta\nu$), the first two $a$-coefficients, structural deformations, and $v \sin i$ (\citealt{gaiadr3}) for the 15 stars we studied in section \ref{sec:deformn}. We also provide the corresponding solar values (\citealt{Solarvsini, Benomar_2023}) for the sake of comparison. While $\Delta\nu$ acts as a proxy for mean stellar density (\citealt{TOUCAN}), the $a_1$ and $a_2$ coefficients measure mean rotation and  structure deformation, respectively.}
    \label{tab:a1a2}
\end{table*}

\subsection{Latitudinal Differential Rotation} \label{sec:ldr}

\begin{figure*}
    \centering
    \includegraphics[width=1\linewidth]{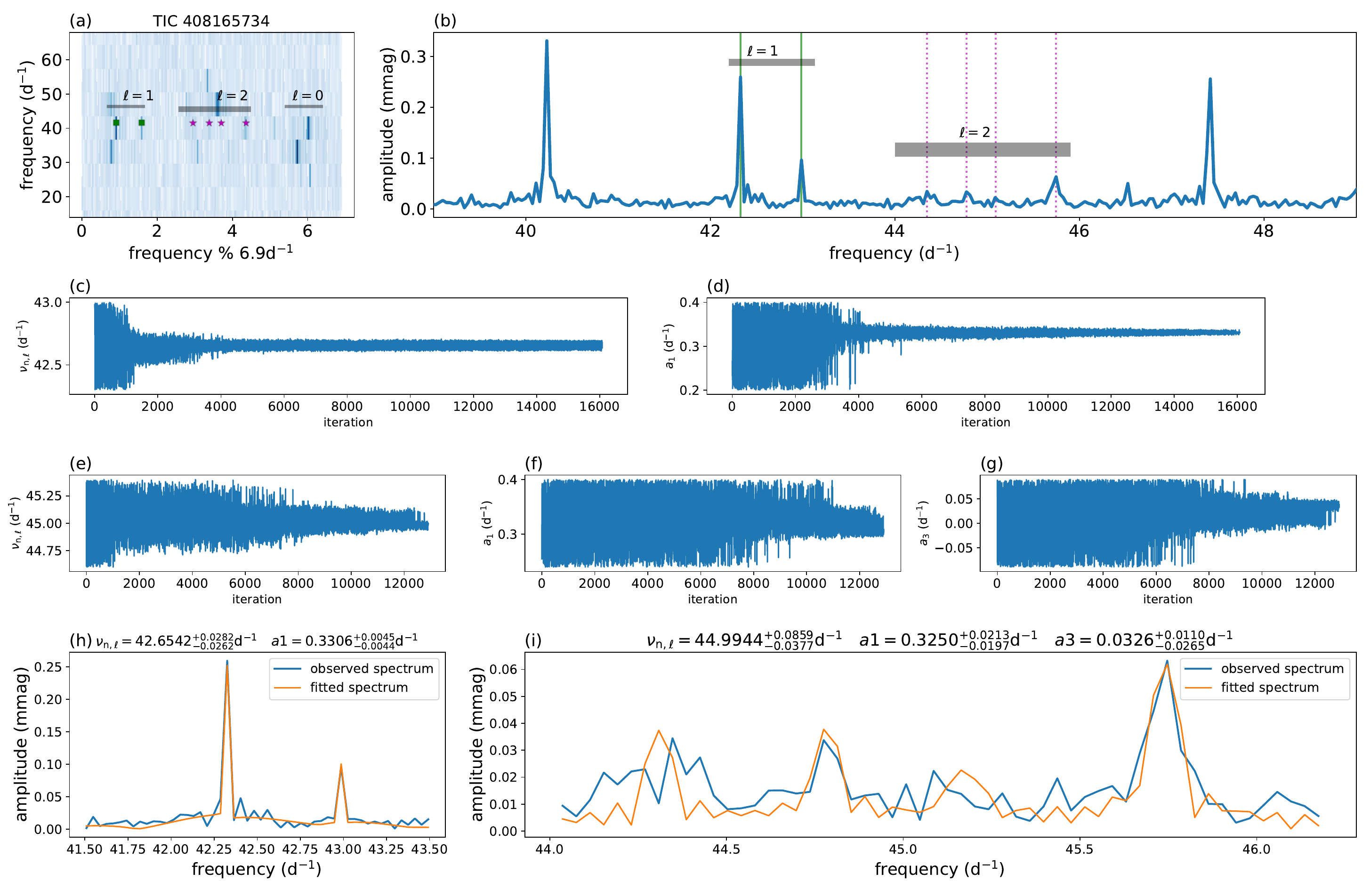}
    \caption{
    Inferring latitudinal differential rotation in a $\delta$ Scuti star.
    (a) \'echelle diagram of TIC 408165734, a star potentially harboring quadrupole ($\ell=2$) modes. The ridge on the right comprises radial ($\ell=0$) modes, and the one on the left appears to be formed by dipole ($\ell=1$) modes. We attribute the multiplet-like structure, midway between these two ridges, to quadrupole modes.
    (b) Oscillation spectrum of the star, with dipole and quadrupole modes marked using solid and dotted lines.
    (c,d) Iterations of parameter $\nu_{\rm n,1}$ and $a_1$, drawn by the sampler to fit the dipole doublet structure in the $41.5 - 43.5 \rm d^{-1}$ frequency range.
    (e,f,g) Iterations of parameter $\nu_{\rm n,2}$, $a_1$, and $a_3$, drawn by the sampler to fit the quadruplet structure in the frequency range of $44.0 - 46.0 \rm d^{-1}$.
    (h,i) Contrasting the observed dipole and quadrupole splittings with spectra modeled with the best fit parameters, shown in figure title, in $\rm d^{-1}$ unit.
    }
    \label{fig:a3_demn}
\end{figure*}

The \'echelle diagrams of $\delta$ Scuti stars often include ridges with modes that are not identified with degrees $\ell=$ 0 or 1. These ridges are speculated to represent higher-degree ($\ell=2$) harmonics (\citealt{Nature2020}) of stellar oscillations. Quadrupole modes form an additional ridge in the \'echelle diagrams of $\delta$ Scuti stars immediately to the left of the ridge formed by the radial modes (\citealt[Fig.2]{accrt_prms}).

We visualized the \'echelle diagrams of the 38 stars in our sample and found in one star (TIC 408165734) the potential presence of $\ell=2$ multiplets, between the radial and dipole ridges (Fig. \ref{fig:a3_demn}a). It comprised four closely spaced modes which we interpreted to be $m=\{ \pm 1, \pm 2 \}$ modes (Fig. \ref{fig:a3_demn}b). The $m=0$ mode is absent in this multiplet, which may be attributed either to a higher inclination angle or a (unknown) mode-suppressing mechanisms. In addition to the presence of quadrupole modes, this star exhibits a dipole doublet (Fig. \ref{fig:a3_demn}b) -- which also lacks the $m=0$ mode, perhaps for similar reason.

Substituting $\ell=2$ in equation \ref{eq:a_cf_exp}, the frequency splittings of quadrupole modes may be expressed in terms of the first four $a$-coefficients:
\begin{align} \label{eq:a_exp_for_l2}
    \nu_{n,2,m} = \nu_{n,2} &+ a_1 m + a_2 (m^2-2) + a_3 \left( \dfrac{5m^3-17m}{3} \right) \nonumber \\
    &+ a_4 \left( \dfrac{35m^4-155m^2+72}{6} \right).
\end{align}
Quadrupole modes are the lowest angular degree vibrations that can be used to infer the $a_3$ coefficient, a quantity that captures latitudinal rotational shear. This coefficient indicates the relative displacement between the centroids of the $m=\pm1$ and $m=\pm2$ components (equation \ref{eq:a3_direct_fm}), and may be calculated from the frequencies of $m=\{\pm 1, \pm 2\}$ multiplets without requiring the $m=0$ component.

The presence of the dipole splitting in this star is very useful in corroborating the information obtained from the quadrupole splittings. Fitting the dipole doublet with equation \ref{eq:a_exp_for_l1}, following the procedure similar as in section \ref{sec:av_rot_splitting}, we obtained $\nu_{\rm n,1} \approx 42.654\pm0.030 ~\rm d^{-1}$ and $a_1 \approx 0.331\pm0.005 ~\rm d^{-1}$.
Similarly, using the equation \ref{eq:a_exp_for_l2}, we attempted to fit the quadrupole splitting, modeled by 13 free parameters: the unperturbed central frequency $\nu_{\rm n,2}$; coefficients $a_1$, $a_2$, $a_3$, and $a_4$; heights ($h_m$) and widths ($w_m$) of the $m=\{\pm1,\pm2\}$ components. Using \texttt{Dynesty}, we sampled these parameters and analyzing the last half iterations, obtained the values of $\nu_{\rm n,2} \approx 44.994_{-0.038}^{+0.086} ~\rm d^{-1}$, $a_1 \approx 0.325\pm0.020 ~\rm d^{-1}$ and $a_3 \approx 0.033_{-0.026}^{+0.011} ~\rm d^{-1}$. The consistency between the values of the $a_1$ coefficients ($\propto$ mean rotation rates) obtained from the independent splittings (i.e. $\ell=$ 1 and 2) suggests that our mode identifications of the quadruplets may be correct.

The positive sign of $a_3$ indicates that latitudinal shear in this star is solar-like, wherein the equator rotates faster than the pole. The ratio $a_3/a_1\sim$10\% in this star, suggesting that rotation in the envelopes of hot stars such as $\delta$ Scuti stars departs from constancy across their meridional cross-sections. In comparison, $a_3$ for the Sun is also approximately 10\% of its mean rotation $a_1$. Self-consistent simulations (\citealt{merid_circ_alphL}) of rotation and meridional circulation in $\alpha$ Leo, a fast-rotating young star, indicate that it ought to show solar-like latitudinal shear and a companion meridional circulation of much smaller magnitude. Meridional flows play a crucial role in maintaining angular momentum balance and a latitudinal differential rotation profile cannot be maintained without this circulation (\citealt{Hanasoge2022}).  

\section{Discussion} \label{sec:Discussion}

In the context of stellar angular momentum transport, \cite{Fuller2019} described how rotational shear in stars can twist and amplify weak poloidal magnetic ($B_{r,\theta}$) fields, transforming them into toroidal ($B_{\phi}$). This would result in the generation of a torque ($\propto B_{r,\theta} B_{\phi}$) that would act to diminish the amplitude of the rotational gradient, in turn reducing the extent of stellar differential rotation. That TIC 307930890 exhibits a strong signature of radial differential rotation in its outermost envelope (figure \ref{fig:rad_df_rot3}d, section \ref{sec:rot_coordinate}) suggests the envelope of this star may be completely non-magnetic ensuring the absence of magnetic torques which would have neutralized the radial rotation gradient. Spectropolarimetric observations of this star can be useful to support or rule out this interpretation. As discussed in the Introduction, $\delta$ Scuti stars display a diverse range of magnetism -- while a dominating toroidal field has been discovered in HD 67523 (\citealt{ds_2_mgntic}), no significant magnetic field was found in a sample of 12 prospective $\delta$ Scuti stars (\citealt{B_dlt_str}).

Six $\delta$ Scuti stars that we examined exhibit asymmetry $a_2>0$, associated with prolate deformation (although at the level of 1-$\sigma$ confidence). Longer time-series observations of these stars may improve the frequency resolution, thus allowing us to determine their prolateness with higher confidence. If they are indeed prolate, equatorial magnetic field is one possible candidate with which to explain such structure. However, a substantially large magnetic field would be required to overcome the oblateness caused by the centrifugal force. \cite{non-sphrl-dtr} performed an order-of-magnitude calculation and found that the acoustic frequency perturbation due to a $10^5$G magnetic field is smaller than the second-order (centrifugal) effect of rotation in a typical $\delta$ Scuti star.

It is possible to determine the subsurface magnetic fields in the 6 prolate stars from their asymmetric frequency splittings. However, it requires sophisticated treatment of the effects of the field, as we elaborate below. Since both rotation and magnetic fields contribute to asymmetric frequency splitting, a non-perturbative calculation is essential to correctly model the effect of rotation, instead of just using the first- and second-order perturbative terms (\citealt{Suarez_2006}) of Coriolis and centrifugal effect, respectively. As the observed modes in our stars are in the non-asymptotic domain ($n \sim 1-4$), formalisms developed for magnetic splitting of asymptotic modes (\citealt{Mathis_2021}) may not be valid for the present purpose. The expression given in \citet{Mathis_2021} is very useful for estimating the total horizontal magnetic field (latitudinal and azimuthal). However, it is not straightforward to measure the strength of the azimuthal (toroidal) field needed to turn the star prolate. Moreover, the magnetic field can influence the acoustic frequencies in two different ways: directly, where the Lorentz force perturbs the pulsation frequencies, and indirectly, where the Lorentz force modifies the equilibrium stellar structure (\citealt{indirect1}) and magnetohydrodynamic pulsation cavities (\citealt{cntrfg_efft}). In the subsurface layers of $\delta$ Scuti stars, the indirect effect is comparable to the direct effect, and hence a simple perturbative calculation will not capture the contribution of the former. Incorporating these ingredients is currently beyond the scope of the present work, although such a complete calculation can provide insight into the magnetic fields of $\delta$ Scuti stars.

{\it Acknowledgments:}
This work has made use of the SIMBAD and VizieR databases. We are thankful for the valuable data released by the NASA's TESS mission and the European Space Agency (ESA) space mission Gaia. We applied \texttt{Lightkurve} (\citealt{Lightkurve}), a Python package for analyzing the Kepler and TESS data. We acknowledge support from the DAE, Government of India (grant no. RTI 4002). All computations were performed in the Intel Lab Academic Compute Environment. We used the \texttt{MESA} and \texttt{GYRE} codes available at Zenodo: doi:\href{https://doi.org/10.5281/zenodo.13879880}{10.5281/zenodo.13879880}. This research was supported in part by a generous donation (from the Murty Trust) aimed at enabling advances in astrophysics through the use of machine learning. Murty Trust, an initiative of the Murty Foundation, is a not-for-profit organisation dedicated to the preservation and celebration of culture, science, and knowledge systems born out of India. The Murty Trust is headed by Mrs. Sudha Murty and Mr. Rohan Murty. We are grateful to the referee for providing constructive comments.

\end{document}